\documentclass[pre,preprint,showpacs]{revtex4}
\usepackage{graphicx,color,epsfig}

\begin{document}

\title{A numerical study of a binary Yukawa model in regimes characteristic
of globular proteins in solution }

\author{Achille Giacometti}
\author{Domenico Gazzillo}
\affiliation{Istituto Nazionale per la Fisica della Materia and \\
Dipartimento di Chimica Fisica, Universit\`{a} di Venezia, \\
S. Marta DD 2137, I-30123 Venezia, Italy}
\author{Giorgio Pastore}
\affiliation{Dipartimento di Fisica Teorica and \\
INFM-DEMOCRITOS, National Simulation Center,\\
Strada Costiera 11, Miramare, I-34100 Trieste, Italy}
\author{Tushar Kanti Das}
\altaffiliation[Present address: ]{Dept. of Physics, Shahjalal University of Science \& Technology, Sylhet-3100
Bangladesh}
\affiliation{ICTP, Diploma Course, Strada Costiera 11, Miramare P.O Box 586, I-34100
Trieste, Italy}

\date{\today }
\begin{abstract}
The main goal of this paper is to assess the limits of validity, in the
regime of low concentration and strong Coulomb coupling (high molecular
charges), for a simple perturbative approximation to the radial distribution
functions (RDF), based upon a low-density expansion of the potential of mean
force and proposed to describe protein-protein interactions in a recent
Small-Angle-Scattering (SAS) experimental study. A highly simplified Yukawa
(screened Coulomb) model of monomers and dimers of a charged globular
protein ($\beta $-lactoglobulin) in solution is considered. We test the
accuracy of the RDF approximation, as a necessary complementary part of the
previous experimental investigation, by comparison with the fluid structure
predicted by approximate integral equations and exact Monte Carlo (MC)
simulations. In the MC calculations, an Ewald construction for Yukawa
potentials has been used to take into account the long-range part 
of the interactions
in the weakly screened cases. Our results confirm that the
perturbative first-order approximation is valid for this system even at
strong Coulomb coupling, provided that the screening is not too weak (i.e.,
for Debye length smaller than monomer radius). A comparison of the MC
results with integral equation calculations shows that both the
hypernetted-chain (HNC) and the Percus-Yevick (PY) closures have a
satisfactory behavior under these regimes, with the HNC being superior
throughout. The relevance of our findings for interpreting SAS results is also discussed.
\end{abstract}


\maketitle


\section{Introduction}

In spite of the large effort devoted in the last decades, a clear
understanding of the interactions of macromolecules in solution is still far
from being achieved \cite{Grier00,Bloor90}. In particular, this is true in
the case of globular proteins, which share with colloidal systems a number
of common properties \cite{Piazza00}.

>From the experimental point of view, there exist several biophysical
techniques for obtaining quantitative data on protein-protein interactions
under physiologically relevant conditions. Small-angle scattering (SAS), for
instance, is currently believed to provide very reliable information, under
very different experimental conditions (pH, ionic strength, temperature,
etc.). If the particle form factors are known, dividing the SAS intensity by
the average form factor yields the experimental average structure factor,
which is related to the radial distribution functions (RDF) $g_{ij}(r)$ ($i$
and $j$ are species indexes). A recent experiment \cite{Spinozzi02} 
reported Small-Angle- X-ray Scattering (SAXS) measurements on
structural properties of a particular globular protein, the $\beta $%
-lactoglobulin ($\beta $LG), in acidic solutions (pH = 2.3), at several
values of ionic strength in the range 7-507 mM. For this protein there is a
clear evidence of a monomer-dimer equilibrium affected by the ionic strength
of the solution \cite{Baldini99}, and the authors of Ref. \cite{Spinozzi02}
were able to achieve a good fit of the experimental data 
by using a highly simplified
\textquotedblleft two-component macroion model\textquotedblright\ (mimicking
monomers and dimers of $\beta $LG), with effective forces represented by
hard-sphere (HS) terms plus the repulsive\ Yukawa (screened Coulomb) part
of the well-known Derjaguin-Landau-Vervey-Overbeek (DLVO) potential \cite%
{Vervey48}. One important novelty of that study, compared with previous
ones, is the proposal of a relatively simple, improved approximation to the
RDFs, suitable for best-fit programs and not restricted to the particular
model but equally well applicable to different spherically-symmetric
potentials.

>From the theoretical point of view, information on intermolecular forces can
be extracted from the experimental average structure factor by comparison
with a theoretical one, whose calculation requires the choice not only of an
interaction model but also of a recipe for deriving the RDFs from the
intermolecular potentials. At present, the most accurate techniques for
evaluating RDFs are the \textquotedblleft exact\textquotedblright\ computer
simulations - Monte Carlo (MC) and molecular dynamics (MD) - and the
\textquotedblleft approximate\textquotedblright\ integral equations (IE)
from the statistical mechanical theory of classical fluids \cite{Hansen86}.
Unfortunately, such complex methods can hardly be included into a best-fit
program for analyzing experimental data. In fact, MC or MD simulations
require long computational times, and become difficult in regimes
characteristic of globular proteins in solution (i.e., low concentration,
high charges, asymmetry in size and charge among the components of the
mixture). On the other hand, only for a very limited number of simple
potentials and within an even more limited number of approximate
\textquotedblleft closures\textquotedblright , IEs of liquid theory admit
analytical solutions, providing closed-form expressions to be inserted into
best-fit codes \cite{Carsughi02}. In all other cases, an iterative numerical
procedure is necessary, and this poses a major drawback to any fitting
scheme. 
Moreover, numerical solution of the IE closures 
tends to become unstable or does
not converge in the region of our interest.

In order to simplify the problem, most analyses of SAS data for highly
dilute solutions employ the crude approximation of neglecting all
intermolecular forces, assuming either large interparticle separations or
weak interactions. In this case, $g_{ij}(r)=1$, the average structure factor
equals unity and the SAS intensity depends only upon the average form
factor. A common first improvement over the previous choice then corresponds
to approximating the RDFs with their zero-density limit, given by the
Boltzmann factor, i.e. $g_{ij}(r)=\exp \left[ -\beta \phi
  _{ij}(r)\right] $, where 
$\phi _{ij}(r)$ is the pair potential, and $\beta =\left( k_{B}T\right)
^{-1} $ the inverse of the thermal energy (absolute temperature multiplied by
Boltzmann's constant). However, this zero-density approximation becomes
insufficient at moderate concentrations or in regimes of colloidal or
protein solutions when electrostatic interactions are strong, i.e. at low
ionic strength.

Motivated by this scenario, Ref. \cite{Spinozzi02} proposed a more accurate
representation of $g_{ij}(r)$ that takes into account, according to a
perturbative scheme, terms up to the first-order in the density expansion of
the potential of mean force, $W_{ij}(r)=-\beta ^{-1}\ln g_{ij}(r)$ \cite%
{Meeron58} (note that the $W_{ij}(r)$-expansion should not be confused with
the RDF one, since these two expansions differ even at the first-order
in density).\ While the satisfactory best-fit results of Ref. \cite%
{Spinozzi02} seem to indicate that a first-order approximation to $W_{ij}(r)$
(W1-approximation) is sufficiently accurate for low concentrations such as
the experimental conditions under study, there is no way, \textit{a priori},
to tell \textit{where} this approximation breaks down, in the absence of
some \textquotedblleft exact results\textquotedblright\ to compare with. On
the other hand, as the experimental conditions present in the analysis of
Ref. \cite{Spinozzi02} are fairly typical in the context of proteins in
solution, we feel that it would be interesting to make such a comparison.
Thus the main subject of the present paper, which complements the
methodological part of the work of Ref. \cite{Spinozzi02}, 
is not the proposal of a
new potential model for $\beta $LG, but a test of the W1-approximation
against more accurate (MC and IE) structural results.

We perform MC simulations, at constant volume, temperature and total number
of macroparticles, for the \textit{same} HS-Yukawa-DLVO binary model,
representing monomers and dimers of $\beta $LG, investigated in Ref. \cite%
{Spinozzi02}. Various values of the screening parameter are considered, and
the MC results for $g_{ij}(r)$ are compared with the corresponding ones
predicted by the aforesaid W1-approximation as well as by some commonly used
IEs. In order
to ensure always a good accuracy, the MC calculations are carried out with
and without a suitable Ewald construction \cite{Kittel76,Rosenfeld96,Salin00}%
, which is expected to play a major role in the cases of strong long-range
interactions (weak screening). Although the theoretical framework for the
Ewald construction, well-known for unscreened Coulomb forces, has already
been extended to Yukawa interactions in recent Refs. \cite%
{Rosenfeld96,Salin00}, this work represents, to the best of our knowledge,
the first MC detailed analysis of its implementation and 
performance for the repulsive Yukawa case \cite{Kahl04}.

Our calculations allow a rather precise determination of the limits of
validity for the W1-expansion. They also show clearly the degree of
reliability of some typical IEs under these frequently encountered,
demanding, regimes. It is worthwhile stressing that our results are in fact
rather general, as there exists a large variety of physical phenomena which
can be described by Yukawa potentials \cite{Rey92}. The existence of
\textquotedblleft exact\textquotedblright\ computer simulations for a binary
model with these potentials would then prove to be useful within a much
more general context than the one treated here.

\section{The protein-protein interaction potential}

When mesoscopic (colloidal or protein) particles with ionizable surface
groups are put into a microscopic polar solvent (like water), most of the
charged surface groups dissociate into the solvent and form microscopic 
\textit{counterions}, usually carrying one or two elementary charges.
Consequently, the big particles acquire high charges of opposite sign and
are called \textit{macroions} or \textit{polyions}. At equilibrium the
counterions are located around the charged macroions, forming an electric
double layer. The counterion distribution tends to screen the repulsions
between macroions, which have charges of the same sign. The result is a
screened Coulomb (Yukawa) repulsion between macroions, which ensures the
stability of the solution (charge-stabilization) with respect to a possible
irreversible flocculation.
An important feature of such repulsions is that they can be
tuned, by adding a suitable amount of a simple electrolyte to the solution.
In fact, such a salt provides additional free \textit{microions} (co-ions,
with same charge sign as the macroions, as well as other counter-ions),
which increase the degree of screening and thus reduce the macroion-macroion
repulsions \cite{Lowen94,Hansen00}.

A $\beta $LG solution thus consists of many components: two different forms
of macroions (protein monomers and dimers), counterions, coions and the
solvent. At neutral pH, the structure of the $\beta $LG protein is dimeric,
while at acidic pH (a condition more similar to the physiological one) a
partial dissociation into two monomers takes place. The monomer-dimer
equilibrium, which determines the molar fractions of both macroion species,
depends upon the ionic strength of the solution. At low ionic strength, the
screening is weak and the electrostatic repulsions predominate over the
attractive forces responsible for the formation of dimers; as a consequence,
most of the macroions are monomers. On the contrary, at high ionic strength
a strong screening reduces the monomer-monomer repulsions in such a way that
a large fraction of dimers can form.

As in Ref. \cite{Spinozzi02}, we represent such a $\beta $LG multicomponent
solution at a highly simplified, \textquotedblleft primitive
model\textquotedblright , level of description, using an effective
\textquotedblleft two-component macroion model\textquotedblright , which
takes into account only protein particles \cite{Lowen94,Nagele96,Hansen00}.
In fact, the solvent is regarded as a uniform dielectric continuum, all
microions are treated as point-like particles, and macroions (both monomers
and dimers) are assumed to be charged hard spheres, with different
diameters. The presence of both solvent and microions appears only in the
macroion-macroion \textit{effective} potentials. In the spirit of the DLVO
theory \cite{Vervey48}, we shall then describe the protein-protein
interactions with the simple effective potential 
\begin{equation}
\phi _{ij}(r)=\phi _{ij}^{\text{HS}}(r)+\phi _{ij}^{\text{Y}}(r)  \label{p1}
\end{equation}%
($i,j=1,2$, with species 1 and 2 corresponding to monomers and dimers,
respectively). Here the hard-sphere term accounts for excluded volume
effects 
\begin{equation}
\phi _{ij}^{\text{HS}}(r)=\left\{ 
\begin{array}{lll}
+\infty , &  & 0<r<\sigma _{ij} \\ 
0, &  & r>\sigma _{ij}%
\end{array}%
\right. ,  \label{p2}
\end{equation}%
where $\sigma _{ij}=(\sigma _{i}+\sigma _{j})/2$ is the distance of closest
approach between two macroparticles of species $i$ and $j$. On the other
hand, the \textit{renormalized }Yukawa term 
\begin{equation}
\phi _{ij}^{\text{Y}}(r)=\frac{Z_{i}Z_{j}e^{2}}{\varepsilon (1+\kappa
_{D}\sigma _{i}/2)(1+\kappa _{D}\sigma _{j}/2)}\ \frac{\exp [-\kappa
_{D}(r-\sigma _{ij})]}{r}  \label{p3}
\end{equation}%
represents an \textit{effective} screened Coulomb repulsion between two 
\textit{isolated} macroions \textit{in a sea of microions}, and has the same
Yukawa form as in the Debye-H\"{u}ckel theory of electrolytes, but with
coupling coefficients of DLVO type \cite{Vervey48}. Here, $e$ is the
elementary charge, $\varepsilon $ the dielectric constant of the solvent, $%
Z_{i}$ the valency of species $i$, and $\kappa _{D}$ is the inverse Debye
screening length due \textit{only }to \textit{microions}, given by 
\begin{equation}
\kappa _{D}=\sqrt{\frac{8\pi \beta e^{2}}{\varepsilon }\frac{N_{A}}{1000}%
\left( I_{c}+I_{s}\right) }.  \label{p4}
\end{equation}%
$N_{A}$ is the Avogadro number, and $I_{c}=(1/2)c_{c}Z_{c}^{2}$ denotes the
ionic strength of the counterions originated from the ionization of the
protein macromolecules (the molar concentration $c_{c}$ of these counterions
is related to the macroion concentrations through the electroneutrality
condition, $c_{c}\left\vert Z_{c}\right\vert =c_{1}\left\vert
Z_{1}\right\vert +c_{2}\left\vert Z_{2}\right\vert $), while $%
I_{s}=(1/2)\sum_{i}c_{i}^{\mathrm{micro}}\left( Z_{i}^{\mathrm{micro}%
}\right) ^{2}$ is the ionic strength of all microions (cations and anions)
generated by added salts. Clearly, $\kappa _{D}^{-1}$ depends on temperature
and represents an indication of the range of the screened Coulomb
interactions, with $\kappa _{D}\rightarrow 0$ corresponding to pure Coulomb
potentials, whereas $\kappa _{D}\rightarrow \infty $ yields the opposite HS
limit. While in real experiments $\kappa _{D}$ is fixed by the chemical
conditions of the solution (namely $I_{c}$ and $I_{s}$), in this work we
shall not consider $I_{s}$ as an independent variable, but, in view
of our methodological purpose, we shall regard $\kappa
_{D}\sigma _{1}\equiv \zeta $ as an independent reduced screening parameter.

A measure of the concentration of the two-macroion effective mixture will be
given by the volume fraction 
\begin{equation}
\eta =\frac{\pi }{6}\sum_{i=1}^{2}\rho _{i}\sigma _{i}^{3}  \label{p5}
\end{equation}%
where $\rho _{i}$ is the partial number density of the $i$-th macroion
species (point-like microions and solvent do not appear here). The
definition of the model is then completed by providing one of the two molar
fractions $x_{i}=\rho _{i}/\rho $ ($i=1,2$), where $\rho =\sum_{i}\rho _{i}$
is the total density.

Now, following partly Ref. \cite{Spinozzi02}, we add three remarks
about some assumptions involved in the choice of the model potential.

i) At first glance one might suspect that reducing dimers to equivalent 
spheres (with a volume twice as large as
the monomer), i.e. neglecting the asymmetry of the dimer molecular shape may
seem a too drastic simplification. In order to clarify this point, it is to
be stressed that in Refs. \cite{Spinozzi02,Baldini99} two different levels
of description for the dimer were used in the two factors which contribute
to the SAS intensity. The coherent scattering intensity $I(q)$ was written
as 
\begin{equation}
I(q)\propto \sum_{i,j}\left( \rho _{i}\rho _{j}\right)
^{1/2}F_{i}^{*}(q)F_{j}(q)S_{ij}(q),  \label{p6}
\end{equation}%
where $q$ is the magnitude of the scattering vector, $F_{i}(q)$ the angular
average of the form factor of species $i$, and the Ashcroft-Langreth partial
structure factors (for spherically-symmetric intermolecular potentials) are
defined by

\begin{equation}
S_{ij}(q)=\delta _{ij}+4\pi \left( \rho _{i}\rho _{j}\right)
^{1/2}\int_{0}^{\infty }r^{2} h_{ij}(r) \frac{\sin \left(
qr\right) }{qr}dr  \label{p7}
\end{equation}%
in terms of the three-dimensional Fourier transform of 
$h_{ij}(r)=g_{ij}(r)-1$. A very
accurate procedure was used to calculate numerically both macroion form
factors, $F_{1}(q)$ and $F_{2}(q)$, from crystallographic data, taking into
account, in particular, the \textit{exact} elongated shape and structure of
the dimer, i.e. its distribution of scattering matter \cite%
{Spinozzi02,Baldini99}. Thus the approximation of spherical dimers was
restricted only to the calculation of $S_{ij}(q)$, which is related, through 
$g_{ij}(r)$, to the intermolecular potentials. At low protein
concentrations, the choice of spherically-symmetric hard-core potentials can
indeed be justified. As in such regimes the average distance 
among particles
is large, intermolecular forces are dominated by the long-range
electrostatic interactions, whereas the details of the short-range
repulsions (i.e. the excluded volume effects) are irrelevant.

ii) Our potentials are purely repulsive. We have not included the attractive
van der Waals part of the DLVO\ potential for charged colloidal suspensions
(the so-called Hamaker term \cite{Vervey48}), as it has already been shown
to be unnecessary for this system in previous work \cite{Spinozzi02}. The
basic reason is that van der Waals attractions may be fully masked by the
electrostatic repulsions when the latter are strong, and are also negligible
for moderately charged particles with a diameter smaller then 50 nm \cite%
{Nagele96}. Moreover, the Hamaker term diverges at contact, so that, to
circumvent this singularity, the inclusion of the attractive term would
require the addition of a Stern layer of counterions (with finite size)
condensed on the macroion surface \cite{Baldini99}.

iii) Given that the specific protein forms dimers, it appears that the $%
\beta $LG necessarily has a short-range monomer-monomer attraction (related
to the surface groups), that causes the aggregation into dimers and
determines the monomer molar fraction $x_{1}$. One expects this attractive
term (possibly including hydrogen bonding) to be rather complex and
non-spherically-symmetric. If such a contribution were clearly
understood and easily tractable, one could start \ from a more fundamental
viewpoint, choosing a model which considers only monomers and includes the
aforementioned attraction into their pair potential. One could then monitor
the dimerization fraction within this \textit{one-component} system.
However, this analysis may be a project on its own right and goes beyond the
aims of the present study.

More simply, in order to avoid poorly known and angular-dependent
potentials, the authors of Refs. \cite{Spinozzi02,Baldini99} adopted 
the viewpoint of using
a \textit{binary }(monomer-dimer) rather than a one-component model, and the
required attraction was accounted only indirectly, by using a chemical
association equilibrium to evaluate $x_{1}$ \cite{Spinozzi02,Baldini99}.

While the dependence of $x_{1}$ upon the added salt (i.e. upon $I_{s}$) must
be taken into account in any best-fit analysis with the binary model \cite%
{Spinozzi02,Baldini99}, in the present work, for the sake of simplicity, we
shall consider $x_{1}$ as an independent parameter. Most of our calculations
will be performed at equal molar fractions, $x_{1}=x_{2}$, but in the last
part of the paper we shall also address the effect of changing the molar
fractions.

\section{Low-density expansion of the mean force potential}

As discussed in the introduction, one of the most commonly used procedures
to compute RDFs $g_{ij}(r)$ for a given pair potential $\phi _{ij}(r)$ goes
through the solution of the Ornstein-Zernike (OZ) IEs from the liquid state
theory, within some approximate closure relation. This can typically be done
only numerically, with the exception of few simple cases (for some
potentials and peculiar closures), where the solution can be worked out
analytically \cite{Hansen86}.

Note that, for HS-Yukawa potentials, the OZ equations do admit analytical
solution \cite{Blum78,Ginoza90,Hayter81}, within the so-called
\textquotedblleft mean spherical approximation\textquotedblright\ (MSA), to
be discussed further below. On the other hand, under the experimental regime
which we are interested in \cite{Spinozzi02}, namely low density and strong
electrostatic repulsions (weak screening), the MSA is well known to display
a serious drawback since RDFs may assume unphysical negative values close to
contact distance $\sigma _{ij}$, for particles $i$ and $j$ which repel each
other. To overcome this shortcoming for repulsive Yukawa models, it would be
possible to utilize an analytical \textquotedblleft rescaled
MSA\textquotedblright\ \cite{Nagele96,Hansen82,Ruiz90} (this possibility
will not be investigated in the present paper) or to resort to different
closures.

In general then, only numerical solutions are feasible, and thus IE
algorithms can hardly be included into best-fit programs for the analysis of
SAS results.

The use of analytical solutions, or simple approximations requiring only a
minor computational effort, is clearly much more advantageous when fitting
experimental data. This can be done by resorting to the following exact,
albeit formal, relation

\begin{equation}
g_{ij}\left( r\right) =\exp \left[ -\beta W_{ij}\left( r\right) \right] ,
\label{e1}
\end{equation}%
\begin{equation}
-\beta W_{ij}\left( r\right) =-\beta \phi _{ij}\left( r\right) +\omega
_{ij}(r),  \label{e2}
\end{equation}%
where $W_{ij}\left( r\right) $ is the potential of mean force, which
includes the direct pair potential $\phi _{ij}\left( r\right) $ as well as $%
-\beta ^{-1}\omega _{ij}(r)$, i.e. the indirect interaction between $i$ and $%
j$ due to their interactions with all remaining macroparticles of the fluid.
In the density expansion of $W_{ij}\left( r\right) $ 
\begin{equation}
-\beta W_{ij}\left( r\right) =\ln g_{ij}\left( r\right) =-\beta \phi
_{ij}\left( r\right) +\omega _{ij}^{(1)}(r)\rho +\omega _{ij}^{(2)}(r)\rho
^{2}+\ldots ,  \label{e3}
\end{equation}

\noindent the \textit{exact} power coefficients $\omega _{ij}^{(k)}(r)$ ($
k=1,2,\ldots $) can be computed by using standard diagrammatic techniques 
\cite{Meeron58}, which yield the results in terms of multi-dimensional
integrals of products of Mayer functions 
\begin{equation}
f_{ij}(\ r)=\exp \left[ -\beta \phi _{ij}\left( r\right) \right] -1.
\label{e4}
\end{equation}

In the zero-density limit, $\omega _{ij}(r)$ vanishes and $g_{ij}\left(
r\right) $ reduces to the Boltzmann factor, i.e. 
\begin{equation}
g_{ij}\left( r\right) =\exp \left[ -\beta \phi _{ij}\left( r\right) \right]
\qquad \text{as~}\rho \rightarrow 0,  \label{e5}
\end{equation}%
which represents a $0^{\mathrm{th}}$-order (W0) approximation, frequently
used in the analysis of experimental scattering data. The W0-approximation
avoids the problem of solving the OZ equations, but is largely inaccurate
except, perhaps, at extremely low densities. We then consider the $1^{%
\mathrm{st}}$-order perturbative correction (W1-approximation) \cite%
{Spinozzi02} 
\begin{equation}
g_{ij}\left( r\right) =\exp \left[ -\beta \phi _{ij}\left( r\right) +\omega
_{ij}^{(1)}(r)\rho \right] .  \label{e6}
\end{equation}%
By construction, this expression is never negative, thus overcoming the
major drawback of MSA. The explicit expression of $\omega _{ij}^{(1)}(r)$
reads 
\begin{equation}
\omega _{ij}^{(1)}(r)=\sum_{k}{x_{k}}\gamma _{ij,k}^{(1)}(r)=\sum_{k}{x_{k}}%
\int \mathrm{d}\mathbf{r}^{\prime }~f_{ik}\left( r^{\prime }\right)
~f_{kj}\left( |\mathbf{r-r}^{\prime }|\right) .  \label{e7}
\end{equation}%
The evaluation of the convolution integral $\gamma _{ij,k}^{(1)}(r)$ is 
most easily carried out in bipolar coordinates. 
After an integration over angle variables 
$\gamma _{ij,k}^{(1)}(r)$ reduces to 
\begin{equation}
\gamma _{ij,k}^{(1)}(r)=\frac{2\pi }{r}\int_{0}^{\infty }dx\ \left[
xf_{ik}\left( x\right) \right] \int_{\left\vert x-r\right\vert }^{x+r}dy\
[yf_{kj}(y)].  \label{e8}
\end{equation}

Of course, the use of the W1-approximation is not restricted to the model
of this paper, and the proposed calculation scheme can be equally well
applied to different spherically-symmetric potentials. While it was shown in
Ref. \cite{Spinozzi02} how this first-order correction largely improves the
fit of experimental scattering data, over the W0-one and under those
experimental conditions, little could be said on the limits of validity of
the W1-approximation with respect to an (hypothetical) exact calculation.
This is the reason why we tackle this task here by a comparison with MC
simulations for a binary HS-Yukawa-DLVO system.

\section{MC simulations and Ewald sum for Yukawa fluids}

The difficulties involved in Monte Carlo calculations dealing with pure
Coulomb potentials are well known \cite{Allen87}. It is now widely
appreciated the usefulness of the so-called Ewald sum for long-range
electrostatic interactions \cite{Allen87,Kittel76}. On the other hand, a
similar construction for Yukawa potentials has appeared in the literature
quite recently \cite{Rosenfeld96,Salin00}. We now briefly recall the
procedure detailed in Refs. \cite{Rosenfeld96,Salin00}. In order to keep
notation as simple as possible, we shall restrict ourselves to 
simple Yukawa potentials,
the extension to our actual potential (Eqs. [\ref{p1}-\ref{p3}]) being obvious.
The basic idea is to start with the total potential energy 
\begin{equation}
U=\frac{1}{2}\sum_{\alpha \beta =1}^{N}q_{\alpha }q_{\beta }\frac{e^{-\kappa
_{D}r_{\alpha \beta }}}{r_{\alpha \beta }},  \label{mc1}
\end{equation}%
where $N$ is the total number of macroparticles, $r_{\alpha \beta }=|\mathbf{%
r}_{\alpha }-\mathbf{r}_{\beta }|$ and $q_{\alpha }=Z_{\alpha }e$, $q_{\beta
}=Z_{\beta }e$ are the charges. This term is then split into a sum of two
contributions, one evaluated in real space, while the other is calculated in
momentum space on wave vectors given by $\mathbf{k}=2\pi \mathbf{n}/V^{1/3}$
($V$ \ is the volume of the system and $\mathbf{n}$ a unit vector of integer
components). To this aim an auxiliary continuous Gaussian charge
distribution 
\begin{equation}
\rho _{q}\left( r\right) =\left( \frac{\lambda ^{2}}{\pi }\right)
^{3/2}e^{-\lambda ^{2}r^{2}}  \label{mc2}
\end{equation}%
is exploited. For $\lambda $ values such that the real space contribution is
limited to particles in the basic simulation cell, the final result reads 
\begin{eqnarray}
U &=&\frac{1}{2}{\sum_{\alpha \beta =1}^{N}}{'}q_{\alpha }q_{\beta }\frac{%
\hbox{erfc}\left( \lambda r_{\alpha \beta }+\kappa _{D}/2\lambda \right)
e^{\kappa _{D}r_{\alpha \beta }}+\hbox{erfc}\left( \lambda r_{\alpha \beta
}-\kappa _{D}/2\lambda \right) e^{-\kappa _{D}r_{\alpha \beta }}}{2 r_{\alpha
\beta }}  \nonumber \\
&+& \sum_{\alpha \beta =1}^{N}\frac{1}{V}\sum_{\mathbf{k}}q_{\alpha }q_{\beta }\frac{4\pi }{%
k^{2}+\kappa _{D}{}^{2}}\exp \left( \frac{-(k^{2}+\kappa _{D}{}^{2})}{4\lambda
^{2}}\right) \cos \left( \mathbf{k}_{\alpha \beta }\cdot \mathbf{r}_{\alpha
\beta }\right)  \label{mc3} \\
&+&\sum_{\alpha }q_{\alpha }^{2}\left[ -\frac{2\lambda }{\sqrt{\pi }}\exp
\left( \frac{-\kappa _{D}{}^{2}}{4\lambda ^{2}}\right) +\kappa _{D}\;%
\hbox{erfc}\left( \frac{\kappa _{D}}{2\lambda }\right) \right] ,  \nonumber
\end{eqnarray}%
where in the first sum we exclude the terms with equal indexes 
and we have introduced the complementary error function 
\begin{equation}
\hbox{erfc}(x)=\frac{2}{\sqrt{\pi }}\int_{x}^{+\infty }dz~e^{-z^{2}}.
\label{mc4}
\end{equation}%
The first two terms in Eq.~(\ref{mc3}) represent the real and momentum space
summations respectively, while the last two contributions refer to the
self-energy \cite{Rosenfeld96,Salin00}. In the limit $\kappa _{D}\rightarrow
0$, the above equation reduces to the Coulomb case \cite{Allen87}, as it
should. Eq.~(\ref{mc3}) contains $\lambda $ as an adjustable parameter, and
we have performed a detailed analysis for its optimal choice, so that the
original potential (\ref{mc1}) is recovered for the range of $\kappa _{D}$
values of interest under the experimental conditions of Ref. \cite%
{Spinozzi02}, without using too many terms in the reciprocal space
summation. Our results indicate $\lambda \sim 6.5/L$ (with $L$ being the
side length of the cubic simulation box) to be the optimal choice, which is
of the same order of magnitude of the one typically used in the Coulomb case.

\section{Integral equations}

Our next task is to test the performance of some IEs under the experimental
conditions of Ref.~\cite{Spinozzi02}. This will strengthen the usefulness of
the W1-approximation, in view of its simplicity compared to a typical IE
calculation for a binary mixture. The OZ IEs of the liquid state theory for $%
p$-component mixtures with spherically-symmetric interactions read \cite%
{Hansen86}

\begin{equation}
h_{ij}\left( r\right) =c_{ij}\left( r\right) +\rho \sum_{l=1}^{p}x_{l}\int 
\mathrm{d}\mathbf{r}^{\prime }~c_{il}\left( r^{\prime }\right)
~h_{lj}\left( |\mathbf{r-r}^{\prime }|\right) ,  \label{ie1}
\end{equation}%
and their solution can be accomplished only in the presence of an additional
approximate relation (closure) between the direct correlation function (DCF) 
$c_{ij}\left( r\right) $ and the total correlation function $h_{ij}\left(
r\right) =g_{ij}\left( r\right) -1$ ($p=2$ in the present case). The most
known among these approximations  are \cite{Hansen86}

\begin{description}
\item {1)} The Percus-Yevick (PY) closure 
\begin{equation}
c_{ij}\left( r\right) =\left[ e^{-\beta \phi _{ij}\left( r\right) }-1\right] %
\left[ 1+\gamma _{ij}\left( r\right) \right] ,  \label{ie2}
\end{equation}

\item where $\gamma _{ij}\left( r\right) =h_{ij}\left( r\right)
-c_{ij}\left( r\right) .$

\item {2)} The Hypernetted Chain (HNC) closure 
\begin{equation}
c_{ij}\left( r\right) =e^{-\beta \phi _{ij}\left( r\right) +\gamma
_{ij}\left( r\right) }-1-\gamma _{ij}\left( r\right)  \label{ie3}
\end{equation}

\item {3)} The Mean Spherical Approximation (MSA), much simpler than the
above two, with the DCF being related only to the potential outside the core 
\begin{equation}
c_{ij}\left( r\right) =-\beta \phi _{ij}\left( r\right) \qquad r\geq \sigma
_{ij},  \label{ie4}
\end{equation}
complemented by the condition of
excluded volume, $g_{ij}(r)=0$ inside the hard cores
\end{description}

Other possible more refined closures, which can be regarded as a combination
of the above three, will be also briefly addressed in this work.

\section{Numerical Results}

A $\beta $LG-monomer is composed of 162 amino acid residues; 20 of these are
basic, so that at pH = 2.3 the monomer is expected to be positively charged,
with about 20 proton charges. In our calculations we fix all
parameters close to their best-fit \textquotedblleft
experimental\textquotedblright\ values \cite{Spinozzi02}, $\sigma
_{1}=40$ \textrm{\AA }, $\sigma _{2}=2^{1/3}\sigma
_{1}\simeq \allowbreak 50.\,\allowbreak 40$ \textrm{\AA\ }, $Z_{1}=20$, $%
Z_{2}=40$, $T = 298.15$ K and $\varepsilon = 78.5$  (strictly
speaking, in Ref. \cite{Spinozzi02} $T = 293.15$,
$\sigma_{1}=38.\,\allowbreak 30$ \textrm{\AA }, and the ratio $%
Z_{2}/Z_{1}$ was about $1.8$, since two of the 20 amino acids of the monomer
are at the monomer-monomer interface in the dimer).

The packing fraction $\eta =0.01$ is also very close to that determined from
the experimental protein concentration $\left( \eta =0.0096\right) $ \cite%
{Spinozzi02}. We then vary the dimensionless screening parameter $\zeta
=\kappa _{D}\sigma _{1}$ in the range $\zeta \sim 1-10$, roughly equivalent
to the range of \ ionic strength $I_{s}$ (from 7 to 507 mM) examined in the
aforesaid SAS measurements for $\beta $LG \cite{Spinozzi02,Baldini99}\ [
where $\zeta =1.41$ when $I_{s}=7$ mM (weak screening, monomer molar
fraction $x_{1\text{ }}=0.85$), and $\zeta =9.08$ when $I_{s}=507$ mM
(strong screening, $x_{1\text{ }}=0.05$) ]. Note that an increase of $\zeta $
has the effect of reducing not only the range of the HS-Yukawa-DLVO
potentials but also their amplitudes, as described by Eq. (\ref{p3}).

In order to obtain the W1-approximation to the RDFs, we have evaluated all
the convolution terms $\gamma _{ij,k}^{(1)}(r)$, given by Eq. (\ref{e8}), at
the grid points $r_{i}=i\Delta r$ ($i=1,\ldots ,500$), with $\Delta r=1
\mathrm{A}$. At each $r_{i}$ value, the double integral 
Eq.(\ref{e8}) has been
carried out numerically, by using the trapezoidal rule for both $x-$ and $y$%
- integration. For the $x$-integration, we have chosen as upper limit the
value $x_{\mathrm{max}}=\max (x_{\mathrm{cut}},\sigma _{2}+r)$, with $x_{%
\mathrm{cut}}=\sigma _{2}+12/\kappa _{D}$, and as grid size $\Delta x=x_{%
\mathrm{cut}}/400$. For the $y$-integration, $\Delta y=\Delta x$.

The MC simulations have been performed at constant $N,V,T$,\ with and
without the Ewald procedure for a correct treatment of the long-range
electrostatic interactions. 
Most 
calculations refer to a total number of
particles $N=216$, divided in monomers and dimers according to the fixed
monomer molar fraction $x_{1}$. Although the sample size may seem 
rather small with
respect to present-day standards, 
one has to take into account that 
the Ewald construction takes a great computational effort with
increasing $N$. In any case, 
we have carried out some additional calculations
with a larger number of particles in order to check for possible
finite-size effects, and found no significant differences in
the results. Hence, we shall use this value of $N$ throughout, with
one exception which will be described later on. The simulation starts from an appropriate
lattice distribution of molecules. We have typically employed $10^{5}$
equilibration steps to eliminate any memory of the initial configuration
artificially introduced into the fluid. Then $5\times 10^{5}$ additional
steps have been used to collect sufficient information for the statistical
averages required to calculate the RDFs.

With the same parameters we have also solved the OZ
integral equations numerically, by means of an
efficient algorithm proposed by Labik \textit{et al.} \cite{Labik85}
employing 1024 grid points, with a mesh size $\Delta
r=0.01\sigma _{1}$, and 20 basis functions. The PY, HNC and MSA closures
have been employed. As expected, the MSA results (not shown in our Figures)
poorly describe the MC data and exhibit the above-mentioned drawbacks of the
MSA closure in regimes with strong coupling at high dilution \cite{Nagele96}%
. We have explicitly checked that other, more sophisticated, approximations,
such as the the Rogers-Young (RY) closure \cite{Rogers84} or the
Zerah-Hansen (HMSA) one \cite{Zerah86}, which attempt to achieve thermodynamic
consistency of compressibility and virial pressures by interpolating between
two of the above closures (PY-HNC and MSA-HNC, respectively), are of no use
here, in such a thermodynamic consistency is never achieved, presumably
because of the combined effect of low densities and strong long-range
repulsions \cite{Pastore02}.

Finally, both MC and IE calculations for $g_{ij}(r)$ have been compared with
the corresponding results from the first-order W1-approximation, with the
aim to assess the limits of validity where the expression given by Eq.~(\ref%
{e6}) can be safely exploited, under conditions typical of proteins in
solution. As further elaborated below, we find that for values $\zeta
\gtrsim 2$ (i.e. $\kappa _{D}^{-1}\lesssim \sigma _{1}/2$) the
W1-approximation well describes the behavior of the RDFs.

When $\zeta $ is large (in the range $\zeta \sim 5-10$) the Yukawa
interactions are strongly screened, and the RDFs essentially reduce to the
typical HS ones, with the first maximum corresponding to the contact
distance $\sigma _{ij}$.

\begin{figure}[tbp]
\centerline{ \epsfxsize=7.0truein \epsfysize=7.0truein
\epsffile{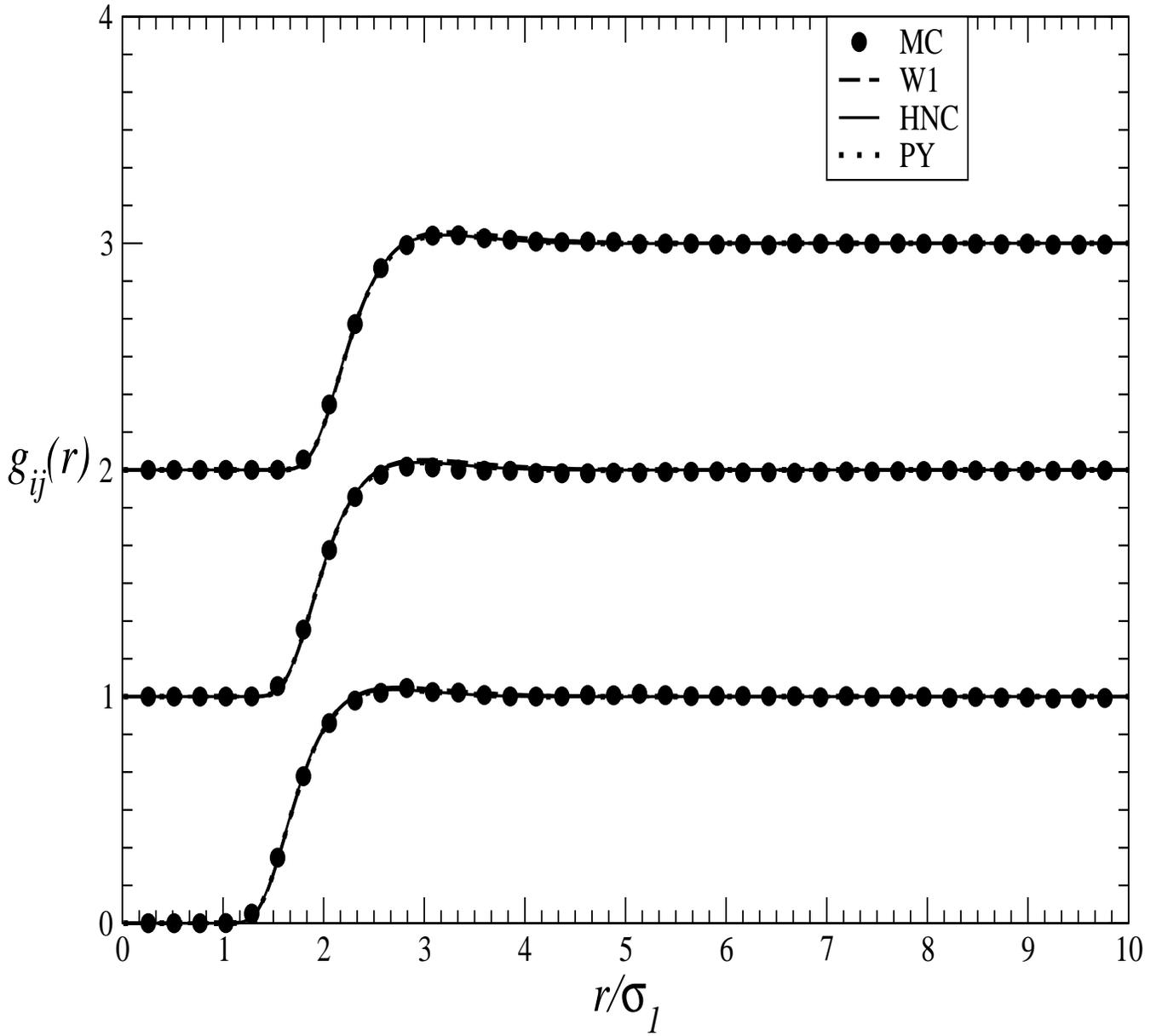} 
}
\vskip1.0cm
\caption{Partial correlation functions $g_{11}(r)$, $g_{12}(r)$ and
  $g_{22}(r)$ (in order from bottom to top) as a function
of the rescaled distance $r/\sigma_1$ for $\zeta=3$ and $x_1=0.5$.
Circles correspond to MC calculations, full lines to HNC, dotted lines to PY, and dashed
lines to the first-order W1-approximation. Here and in the following the
components $12$ and $22$ have been shifted upwards by one and two units, respectively.}
\label{Fig1}
\end{figure}

Fig. \ref{Fig1} depicts the comparison between 
the MC results and the W1-approximation
for $\zeta =3$ (corresponding to a moderately weak screening) and $x_{1}=0.5$%
, that is when both monomers and dimers are present in equal measure. Note
that these conditions are close to one of the experimental cases reported in
Ref. \cite{Spinozzi02}, where $I_{s}=47$ mM corresponds to $\zeta =2.8$ and $%
x_{1}=0.48$. On the other hand, as the ionic strength $I_{s}$ is lowered
from 507 mM to 7 mM, the experimental system switches from a fluid almost
completely made up of dimers ($x_{1}=0.05 $) to one almost completely made
up of monomers ($x_{1}=0.85$). This rather peculiar feature is specific of
the $\beta $LG and will also be considered further on. Here, however, our
main aim is to test the W1-approximation under the simple, symmetric,
condition of equal molar fractions, since we already know, from Ref. \cite%
{Spinozzi02}, that the first-order approximation well describes the $\beta $%
LG experimental data, which display, in particular, a lowering in the
scattering intensity at small angles, with a progressive development of an
interference peak at low ionic strengths. In Fig. \ref{Fig1} we also report
the results from the HNC and PY IEs (solid and dotted lines), which are
practically indistinguishable on the employed scale. It is apparent that in
the case of Fig. \ref{Fig1} the W1-RDFs $g_{11}(r)$, $g_{12}(r)$ 
and $g_{22}(r)$ are
in excellent agreement with their MC, HNC and PY counterparts. Note that, for all
three RDFs, $g_{ij}(r)$ remains zero even in a region outside the hard-core,
while the position of the peak lies at a distance larger than $\sigma _{ij}$%
, as a consequence of the strong Yukawa repulsions.

\begin{figure}[tbp]
\centerline{ \epsfxsize=7.0truein \epsfysize=7.0truein
\epsffile{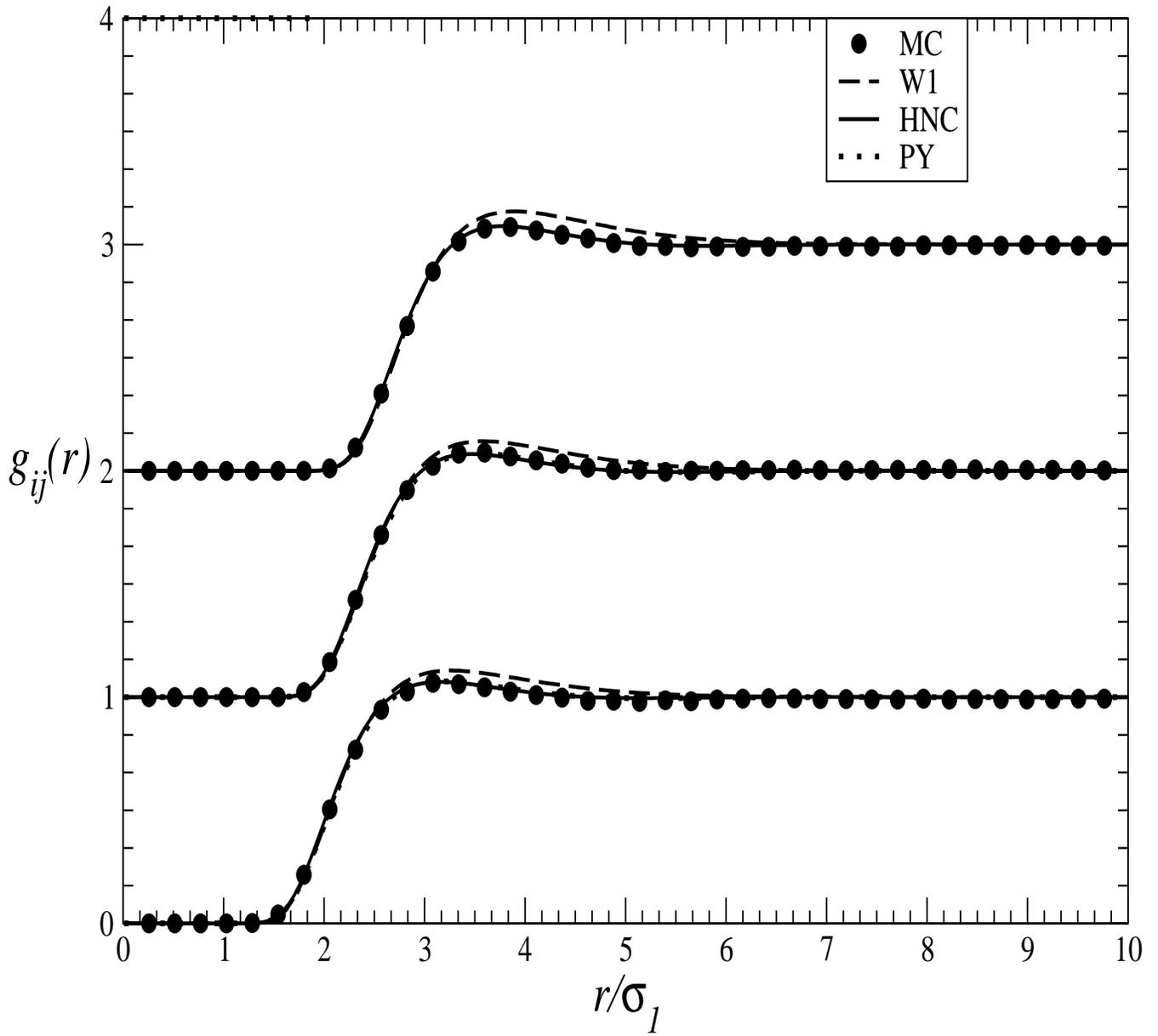} 
}
\vskip1.0cm
\caption{Same as above with $\zeta=2$ and $x_1=0.5$.}
\label{Fig2}
\end{figure}

A departure of the first-order W1-approximation from the MC results
can be observed for smaller values of the screening parameter $\zeta $,
where higher-order terms in the density expansion of $W_{ij}\left( r\right) $%
, Eq. (\ref{e3}), begin to have a non-negligible effect. This is indicated
in Fig. \ref{Fig2} for the case $\zeta =2$, which corresponds to $\kappa
_{D}^{-1} = \sigma _{1}/2$, with the Debye screening length being
equal to the monomer radius (among the experimental data of Ref. \cite%
{Spinozzi02} we find $\zeta =2$ and $x_{1}=0.73$ when $I_{s}=17$ mM). Again
the HNC and PY RDFs are nearly identical with each other and with MC data.
On the other hand, the W1-approximation
predicts peaks nearly at the same positions as the PY and HNC closures,
while its peak heights are slightly overestimated. However, the agreement
between W1 and MC results can still be regarded as rather good.

\begin{figure}[tbp]
\centerline{ \epsfxsize=7.0truein \epsfysize=7.0truein
\epsffile{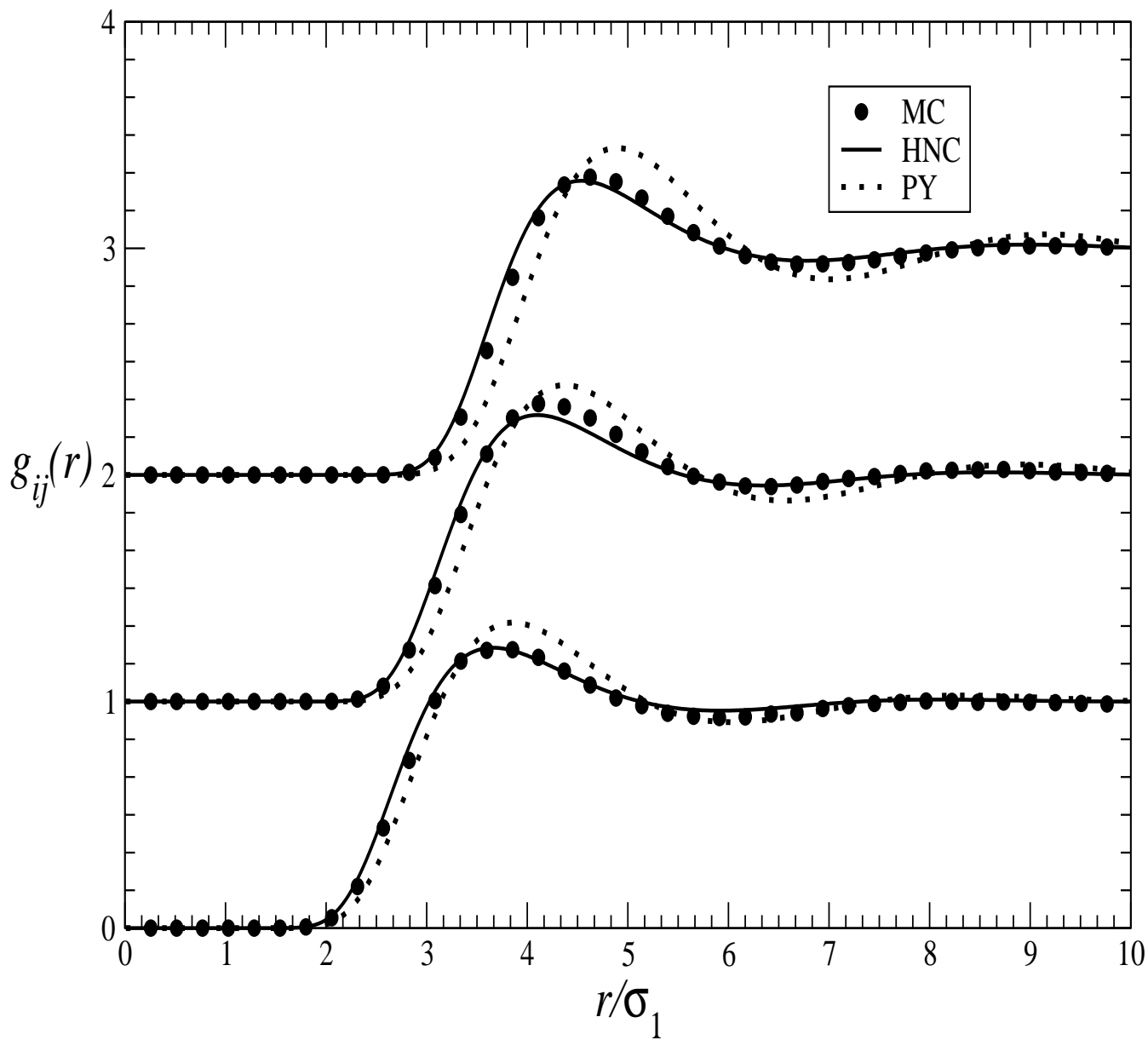} 
}
\vskip1.0cm
\caption{Comparison of the results from HNC, PY and MC
in the calculation of the partial
radial distributions functions for $\zeta=1$ and $x_1=0.5$. The first-order
W1-approximation is not depicted as it overshoots the MC results roughly
by a factor $2$.}
\label{Fig3}
\end{figure}

In regimes with weaker screening the discrepancies become more and more
pronounced. The breakdown of all the considered approximations can be
clearly appreciated in Fig. \ref{Fig3} for $\zeta =1$ 
(note that the case with the
weakest screening in Ref. \cite{Spinozzi02} corresponds to $I_{s}=7$ mM, $%
\zeta =1.41$ and $x_{1}=0.85$). The W1-results are not reported in
this Figure, since they are way off from the MC data (with an
overestimation of about a factor $2$). On the other hand,
even the results from the PY approximation are
significantly displaced from the MC RDFs. The difference between the HNC and
PY results is apparent, particularly for the latter, as expected. 
The PY approximation overestimates both the heights and
positions of the peaks, compared to the HNC ones. Overall the PY
approximation fails to describe the MC calculation for $\zeta <2$,
whereas the HNC closure is consistently in good agreement with the MC data.
Such a good performance of the HNC closure closely resembles
the good agreement between HNC and MC, even
at strong Coulomb coupling, for the \textit{one-component} fluid of \textit{%
point} charges (electron gas or plasma, with $\zeta =0$) (OCP) in a uniform
neutralizing background \cite{Ng74}. However, the results for our
binary model with screening at packing fraction $\eta =0.01$ can hardly be
compared with the available MC simulations for \textit{one-component }%
charged hard spheres (OCCS, with $\zeta =0$) in a uniform neutralizing
background, at $\eta =0.3\div 0.4$ \cite{Hansen77}. Moreover, it is known
the inadequacy of the HNC for high charges at low concentrations (for
instance, in the dilute regime of 2-2 aqueous electrolytes \cite%
{Rossky80,Ciccariello82}, where bridge diagrams become non-negligible for
like-charge RDFs). On the other hand, despite the large number of
comparisons among PY, HNC and MC predictions carried out over the years, we
are not aware of a similar detailed RDF investigation under regimes
characteristic of globular proteins in solution, for HS-Yukawa-DLVO binary
models.

\begin{figure}[tbp]
\centerline{ \epsfxsize=7.0truein \epsfysize=7.0truein
\epsffile{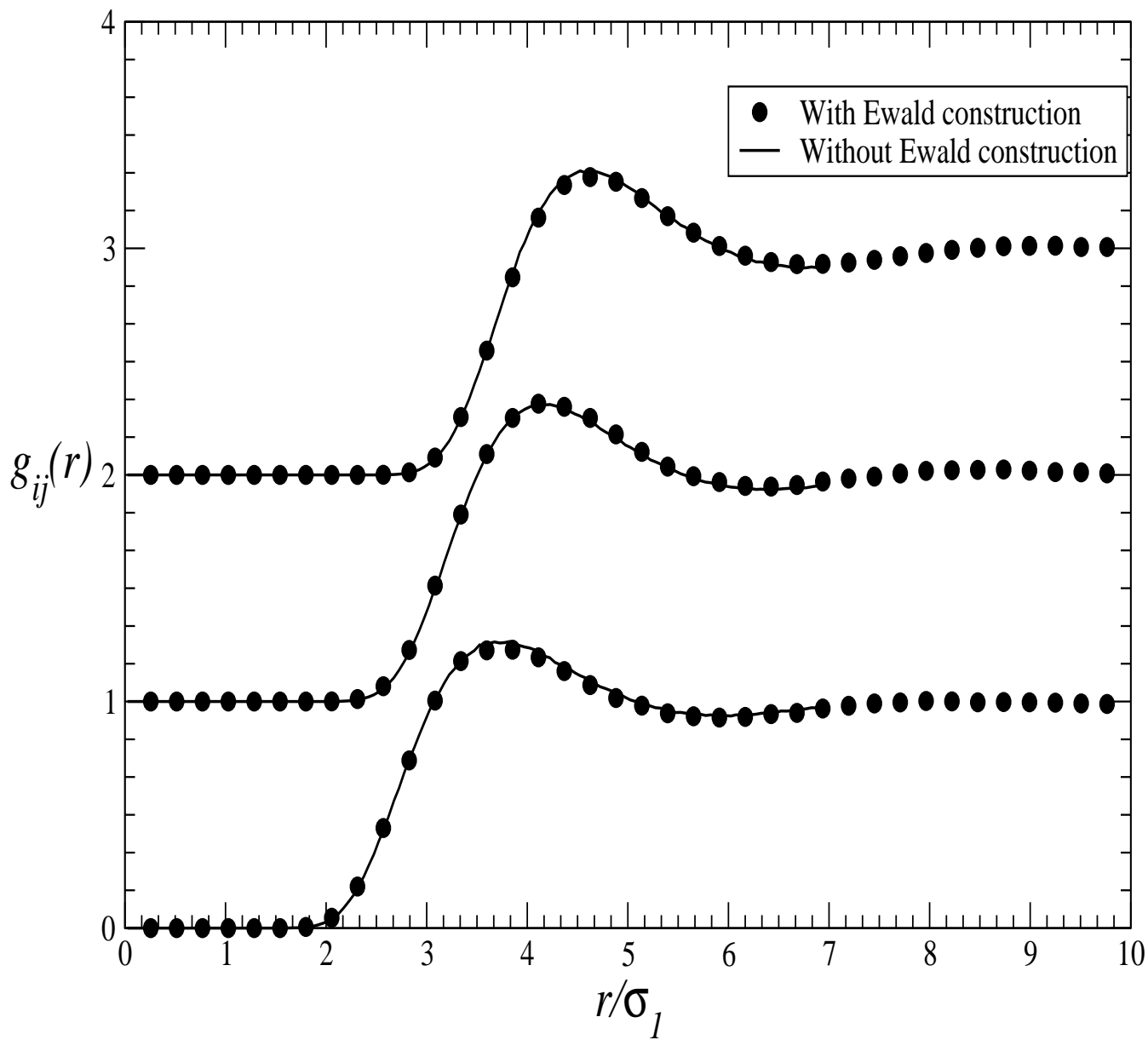} 
}
\vskip1.0cm
\caption{Partial correlation functions $g_{11}(r)$, $g_{12}(r)$ and
  $g_{22}(r)$ (in order from bottom to top) as a function
of the rescaled distance $r/\sigma_1$,  as computed with (circles) and without
(solid line) the Edwald construction, for $\zeta=1$ and $x_1=0.5$.}
\label{Fig4}
\end{figure}

\begin{figure}[tbp]
\centerline{ \epsfxsize=7.0truein \epsfysize=7.0truein
\epsffile{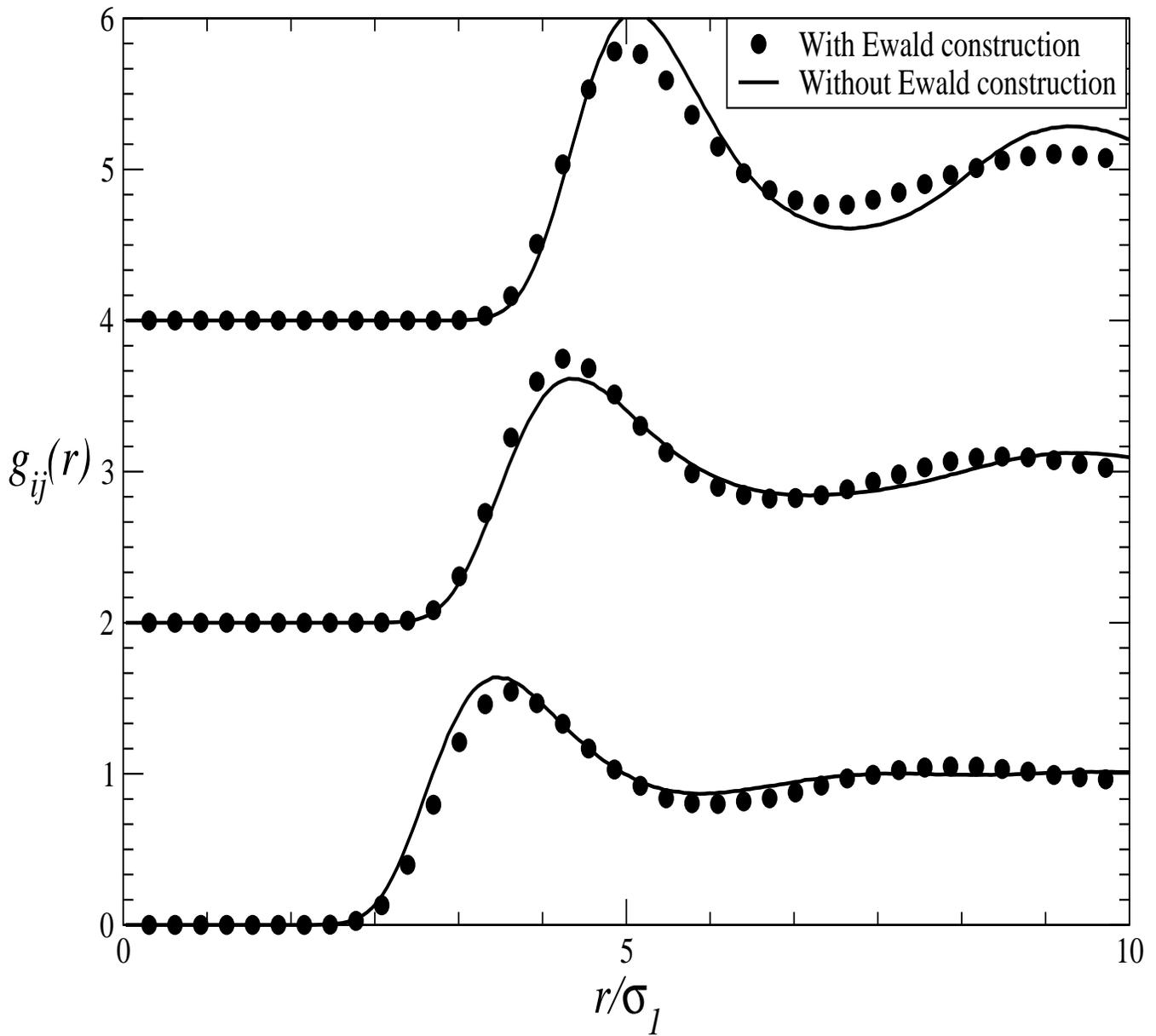} 
}
\vskip1.0cm
\caption{Same as above with $\zeta=0.25$ and $x_1=0.5$. Note that the
  scale has been changed with respect to previous figures. Accordingly, 
here components $12$ and $22$ have been shifted upward by $2$ and $4$ units,
respectively.}
\label{Fig5}
\end{figure}

Next we consider the effect of taking into proper account 
the long-range nature of the interactions (in the weakly screened case)
with the use of the Ewald construction. This is illustrated in Figs. \ref{Fig4}
and \ref{Fig5}, where the RDFs computed with and without
the Ewald construction are compared
at $\zeta =1$ and $\zeta =0.25$, respectively.
Clearly, very little difference is detected between these two calculations
when $\zeta =1$ (and when $\zeta=0.5$, not shown). We find that the the presence of the Ewald construction
begins to be important for very low values of the screening parameter ($%
\zeta \lesssim 0.25$, i.e. $\kappa _{D}^{-1}\gtrsim 4\sigma _{1}$), as shown
in Fig. \ref{Fig5}. Supplementary calculations, not reported here, 
confirm that this
is true even for lower values of protein charges, that is for weaker Coulomb
coupling.

\begin{figure}[tbp]
\centerline{ \epsfxsize=7.0truein \epsfysize=7.0truein
\epsffile{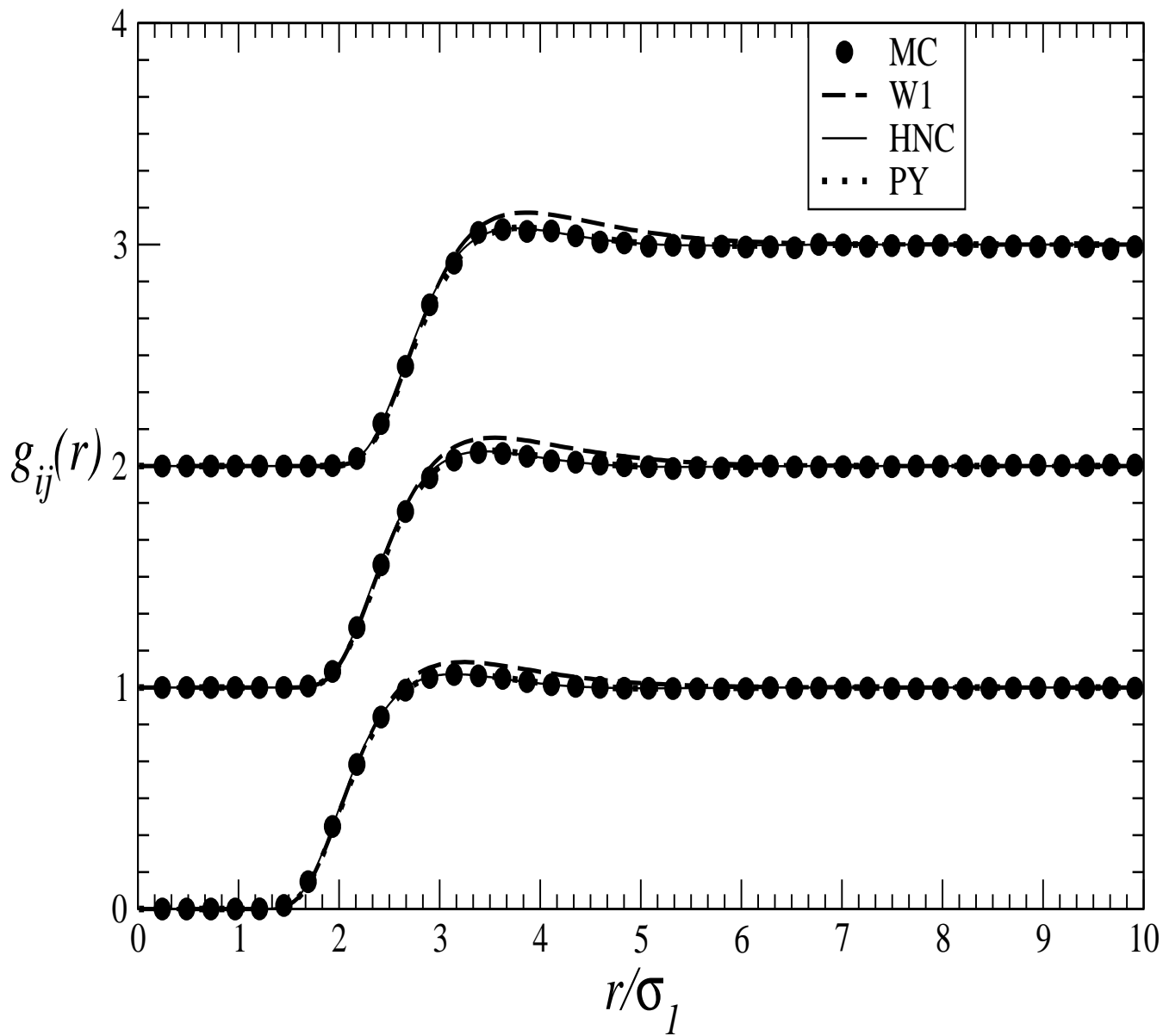} 
}
\vskip1.0cm
\caption{Asymmetric case, corresponding to Fig. \ref{Fig2}, with
  $\zeta=2$ and $x_1=0.75$.}
\label{Fig6}
\end{figure}

Finally, we consider the effect of varying the molar fractions. While the
exact conditions reported in the $\beta $LG experiment pose a very difficult
challenge to an accurate MC calculation in view of the particular
combination of strong asymmetry and repulsions, we can nevertheless easily
account for the general trend. This is depicted in Fig. \ref{Fig6}, 
where we have
assumed $\zeta =2$ and $x_{1}=0.75$, in closer analogy with a $\beta $LG
experimental case, $\zeta =2$ and $x_{1}=0.73$ when $I_{s}=17$ mM. It is
apparent how the performance of the first-order W1-approximation is comparable
to the corresponding symmetric case, $\zeta =2$ and $x_{1}=0.5$.

\begin{figure}[tbp]
\centerline{ \epsfxsize=7.0truein \epsfysize=7.0truein
\epsffile{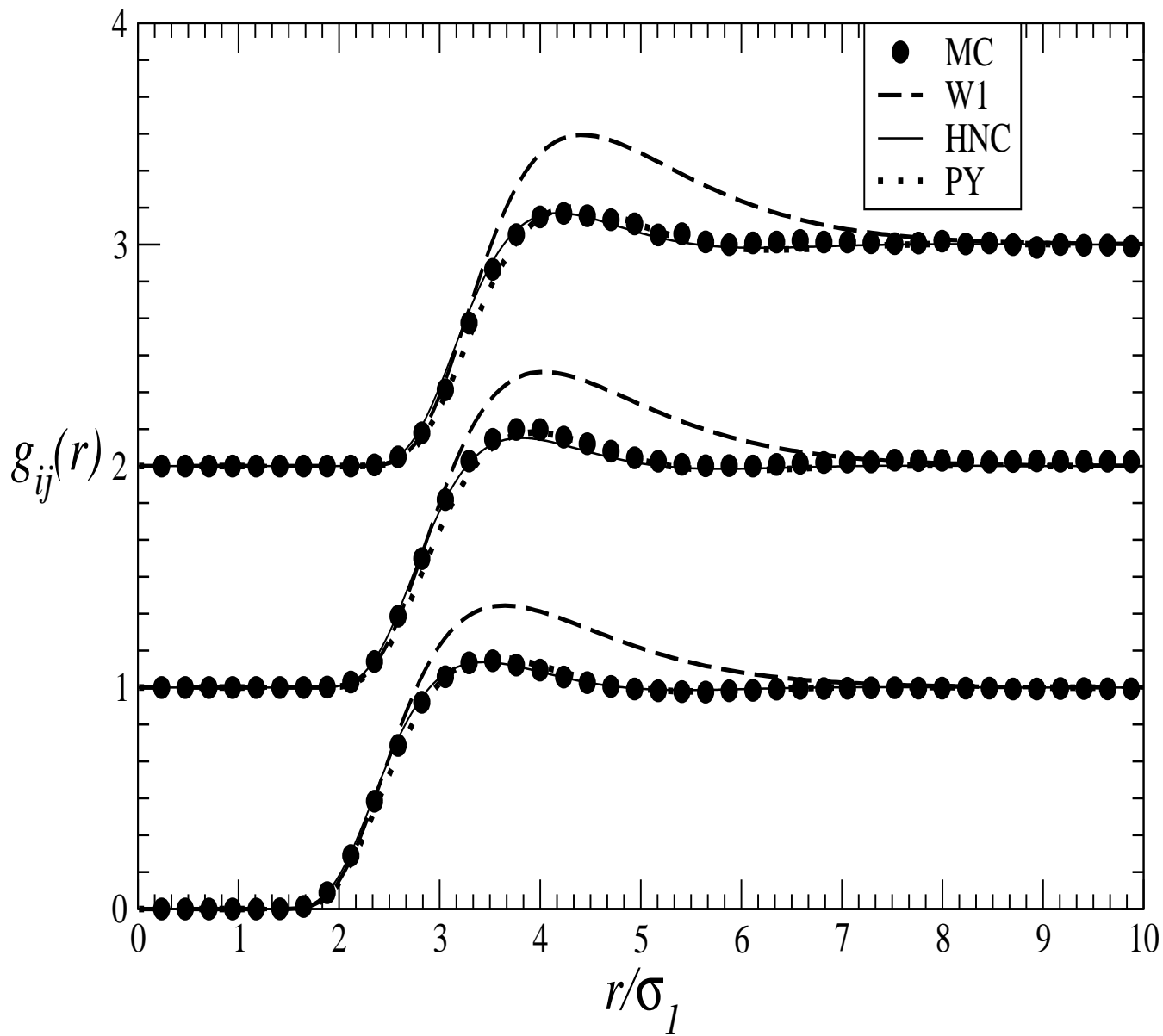} 
}
\vskip1.0cm
\caption{Asymmetric case with $\zeta=1.41$, $x_1=0.85$,
  corresponding to the lowest value of ionic strength $I_{s}=7$ mM (i.e., the
  weakest screening) investigated in Ref. \cite{Spinozzi02}.}
\label{Fig7}
\end{figure}

Fig. \ref{Fig7} refers to the asymmetric case with the 
weakest screening in Ref. 
\cite{Spinozzi02}, i.e. $\zeta =1.41$ and
 $x_{1}=0.85$ (corresponding to the lowest value of
ionic strength, $I_{s}=7$ mM). Again, the HNC and PY results
are in good agreement with the MC ones, and even the performance of
the W1-approximation can be regarded as acceptable, in
agreement with the results of Ref. \cite{Spinozzi02}. We note
that, in view of the low molar fraction of species $2$ (dimers), the
results of Fig. \ref{Fig7} refer to a higher number of particles
($N=512$). 

\section{Conclusive remarks}

This work represents a necessary 
verification
of the best-fit analysis of SAS
experimental data, for solutions of $\beta $-lactoglobulin, presented in
Ref. \cite{Spinozzi02}. In the present paper we have assessed the
limits of validity of the W1-approximation, exploited in that work
to calculate, in a simple way, the RDFs in regimes typical of a large class
of \ globular proteins in solution, that is low concentrations and high
macroion charges. This task has been accomplished by considering the \textit{%
same} highly simplified model proposed in Ref. \cite{Spinozzi02} (i.e. a 
\textit{binary} mixture of monomers and dimers of the protein, with
HS-Yukawa-DLVO effective potentials), and comparing the corresponding $%
g_{ij}(r)$ obtained by three different methods: the first-order density
expansion of the potential of mean force (W1-approximation),
\textquotedblleft exact\textquotedblright\ MC simulations and approximate
IEs. All results reported here refer to $\eta =0.01$ and high macroion
charges, $Z_{1}=20$ and $Z_{2}=40$. For the MC simulations we have
implemented an Ewald construction for Yukawa potentials, which ensures a
proper treatment of the long-range part of the interactions, and we have
tested its relevance as a function of the screening parameter $\zeta $. In
the IE calculations simple closures (PY, HNC, and MSA) as well as more
elaborated ones (RY and HMSA) have been considered.

We can summarize the obtained results as follows.

\begin{description}
\item {i)} The first-order W1-approximation can be considered reliable in
regimes with low concentration ($\eta =0.01$) even for strong Coulomb
coupling (up to charges of $10\div 20e$ on macroions with diameters of $%
40\div 50$ \textrm{\AA }), provided that the screening is strong enough,
i.e. when $\zeta \gtrsim 2$ or, equivalently, $\kappa _{D}^{-1}\lesssim
\sigma _{1}/2$ (Debye length smaller than monomer radius). This finding
demonstrates that the previous usage of the W1-approximation in Ref. \cite%
{Spinozzi02} was fully legitimate, for all considered cases including those
with the lowest ionic strength ($I_{s}=7$ mM, $x_{1}=0.85$, $\zeta
=1.41$), which lies near the borderline of the reliability region. For
weaker screening (lower values of  $\zeta $ or larger $\kappa _{D}^{-1}$) at
least second-order terms in the density expansion should be taken into
account. However, the resulting W2-approximation would require a much higher
computational effort and thus could not be conveniently included into a
best-fit program for analyzing SAS experimental data.

\item {ii)} In the MC simulations the Ewald construction for Yukawa
potentials starts to be important for weak screening corresponding to $\zeta
\lesssim 0.25$ ($\kappa _{D}^{-1}\gtrsim 4\sigma _{1}$), and this is true
even for lower values of the protein charges.

\item {iii)} Both the HNC and PY IEs yield sufficiently accurate values of
the RDFs, as long as $\zeta \gtrsim 2$. For lower values of $\zeta $
HNC is still accurate, whereas PY starts to deviate as expected.
The MSA predictions, on the other hand, are
very poor even in those regimes where the W1-approximation can be considered
reliable. Under these conditions both the RY and HMSA closures are found not
to achieve thermodynamic consistency between compressibility and virial
pressures.

\item {iv)} The sufficient accuracy of the W1-approximation in the regimes
of our interest (tested in this paper against \textquotedblleft
exact\textquotedblright\ MC results), together with its success (shown in
Ref. \cite{Spinozzi02}) in reproducing the main features of the experimental
SAS intensity curves for the examined $\beta $LG solutions, confirm the good
performance of the highly idealized two-macroion model, which includes
spherically-symmetric HS-Yukawa-DLVO repulsions, a monomer-dimer chemical
equilibrium, and the \textquotedblleft exact\textquotedblright\ form
factors, evaluated by taking into account the real non-spherical structure
of the dimer.

\end{description}

Clearly, all complex characteristics of the interactions between globular
proteins cannot be explained by the \textquotedblleft
primitive\textquotedblright\ level of description adopted in Ref. \cite%
{Spinozzi02} and here. We have followed the generally accepted philosophy of
exploiting the simplest possible description of the system, which yet can
provide useful information on the basic underlying interaction
mechanism. The
determination of the \textquotedblleft true\textquotedblright\
protein-protein potentials thus remains an open problem. 

Our choice of  \textit{purely repulsive} interactions illustrates the
minimal assumptions allowing a satisfactory reproduction of the SAS data for 
$\beta $LG. In many studies on colloidal or protein solutions, satisfactory
results were obtained from very simplified models. The use of sophisticated
potentials, with a large number of different contributions, is often
unnecessary at the first stages. Moreover, a high level of
description for potentials would be in striking contrast with the poor level
of approximation to the RDFs (W0-approximation) commonly adopted in many
analyses of experimental data.

As regards the approximation of \textit{spherical symmetry}, used for the
protein-protein interactions (but not in the calculation of the form
factors), we remark that it represent a common simplifying choice. In
particular, it is worth recalling a very recent study by Pellicane \textit{%
et al. }\cite{Pellicane04}, which reports evidence that  the phase diagram
of prototype globular protein solutions (lysozyme and $\gamma $-crystallin
in water and added salt) can be reasonably reproduced by a
spherically-symmetric representation of macromolecular interactions. These
authors employed a HS-Yukawa-DLVO one-component potential, including the
Hamaker attractive part.

Evidently, in addition to the molecular granularity of the solvent and the
finite sizes of all microions, a highly refined model description of protein
solutions should embody the asymmetry of the molecular shape as well as the
heterogeneity of the macroion surface charge distribution. The presence of
different charged surface groups may produce \textquotedblleft charge
patches\textquotedblright\ that have a sign opposite to that of the net
macroion charge. The importance of non-spherically-symmetric models with an
inhomogeneous distribution of positively and negatively charged groups was
recently investigated in a MC study on the electrostatic complexation of
flexible polyelectrolytes with $\alpha $-lactalbumin and $\beta $%
-lactoglobulin \cite{deVries04}.

As a final remark to the present paper, it is worth pointing out that
 we are not aware of
any previous investigations of this type within the HS-Yukawa-DLVO binary
model and in regimes typical of globular proteins in solution. Our
results and the methodological approach based upon the W1-approximation are
expected to be useful in the analysis of SAS experiments. It would be rather
interesting to pursue a similar study on the thermodynamic predictions of
the first-order approximation. This could be easily carried out, as all
thermodynamic quantities can be inferred either directly or through the
knowledge of the RDFs. Another interesting issue, within the present
framework, involves an increase of the asymmetry between the two considered
molecular sizes, which is known to lead to possible depletion effects 
\cite{Likos01}. We plan to perform such investigations in a future publication.

\acknowledgments We are particularly grateful to Francesco Spinozzi, Flavio
Carsughi and Paolo Mariani, for enlightening discussions and on-going
collaboration on the subject reported in this work. TKD thanks Prof. S.R.
Shenoy and the Abdus Salam ICTP Trieste for some support. The Italian MIUR
(Ministero dell'Istruzione, dell'Universit\`{a} e della Ricerca)
through a PRIN-COFIN project, and the
INFM (Istituto Nazionale di Fisica della Materia) are gratefully
acknowledged for partial financial support.

\newpage 

\newpage

\newpage
\newpage
\newpage
\newpage


\begin{thebibliography}{910}

\bibitem{Grier00} D.~G.~Grier, E.~R.~Dufresne and S.~H.~Behrens, \textit{
Interactions in colloidal suspensions} (The University of Chicago, USA 2000). 

\bibitem{Bloor90} D.~M.~Bloor and E. Wyn-Jones, \textit{The Structure,
Dynamics and Equilibrium Properties of Colloidal Systems} (Kluver Academic
Publisher, Netherland 1990). 

\bibitem{Piazza00} R. Piazza, Current Opinion in Colloid \& Interface
Science, \textbf{5}, 38 (2000). 

\bibitem{Spinozzi02} F. Spinozzi, D. Gazzillo, A. Giacometti, P. Mariani,
and F. Carsughi, Biophys. J. \textbf{82}, 2165 (2002). 

\bibitem{Baldini99} G. Baldini, S. Beretta, G. Chirico, H. Franz, E.
Maccioni, P. Mariani, and F. Spinozzi, Macromolecules \textbf{32}, 6128
(1999).

\bibitem{Vervey48} E. J.~Vervey, and J. Th. G. Overbeek, \textit{Theory of
the Stability of Lyophobic Colloids}, (Elsevier, Amsterdam. 1948). 

\bibitem{Hansen86} J. P.~Hansen , and I. R. Mc Donald, \textit{The Theory of
Simple Liquids}, (Academic Press, London 1986). 

\bibitem{Carsughi02} F. Carsughi, A. Giacometti and D. Gazzillo, Comp. Phys.
Comm. \textbf{133}, 66 (2002); D. Gazzillo and A. Giacometti, J. Chem. Phys. 
\textbf{113}, 9837 (2000); D. Gazzillo and A. Giacometti, J. Chem. Phys. 
\textbf{120}, 4742 (2004). 

\bibitem{Meeron58} E. Meeron, J. Chem. Phys. \textbf{28}, 630 (1958). 

\bibitem{Kittel76} C. Kittel, \textit{Introduction to Solid State Physics}
(Wiley, New York 1976). 

\bibitem{Rosenfeld96} Y. Rosenfeld, Mol. Phys. \textbf{88}, 1357 (1996). 

\bibitem{Salin00} G. Salin and J. M. Caillol, J. Chem. Phys. \textbf{113},
10459 (2000).

\bibitem{Kahl04} In a recent paper (E. Sch\"{o}ll-Paschinger, D. Levesque, 
J.J. Weiss, and G. Kahl, J. Chem. Phys. \textbf{122}, 024507 (2004)), the same
 Ewald construction has been used in a different regime for {\em attractive}
 hard-sphere Yukawa mixtures.

\bibitem{Rey92} C. Rey, L.~J.~Gallego, L.~E.~Gonz\'{a}lez and D.~J.~Gonz\'{a}%
lez, J. Chem. Phys. \textbf{97}, 5121 (1992).

\bibitem{Lowen94} H. L\"{o}wen, Phys. Reports \textbf{237}, 249 (1994).

\bibitem{Nagele96} G. N\"{a}gele, Phys. Reports \textbf{272}, 215 (1996).

\bibitem{Hansen00} J. P. Hansen, and H. L\"{o}wen, Ann. Rev. Phys. Chem. 
\textbf{51}, 209 (2000).

\bibitem{Blum78} L. Blum, and J.S. Hoye, J. Stat. Phys. \textbf{19,} 317
(1978).

\bibitem{Ginoza90} M. Ginoza, Mol. Phys. \textbf{71}, 145 (1990).

\bibitem{Hayter81} J. B. Hayter, and J. Penfold, Mol. Phys. \textbf{\ 42},
109 (1981).

\bibitem{Hansen82} J. P. Hansen, and J. B. Hayter, Mol. Phys. \textbf{\ 46},
651 (1982).

\bibitem{Ruiz90} H. Ruiz-Estrada, M. Medina-Noyola, and G. N\"{a}gele,
Physica A \textbf{168}, 919 (1990).

\bibitem{Allen87} M. P. Allen and D. J. Tildesley, \textit{Computer Simulation
of Liquids}, (Academic Press, New York 1987).

\bibitem{Labik85} S. Labik, A. Malijevsky, and P. Vonka, Molec. Phys. 
\textbf{56}, 709 (1985).

\bibitem{Rogers84} F. J.~Rogers, and D. A. Young, Phys. Rev. A \textbf{30},
999 (1984).

\bibitem{Zerah86} G. Zerah, and J. P. Hansen, J. Chem. Phys. \textbf{84} 2336
(1986).

\bibitem{Pastore02} Note that the possibility of a lack of thermodynamic
consistency have been already noted earlier both for RY (G. N\"{a}gele,
Phys. Reports \textbf{272}, 215 (1996)) and HMSA (G. Kahl and G. Pastore,
Europhys. Lett. \textbf{7}, 37 (1988); F. Ould-Kaddour and G. Pastore, Mol.
Phys. \textbf{81}, 1011 (1994).

\bibitem{Ng74} K.\ G. Ng, J. Chem. Phys. \textbf{61} 2680 (1974).

\bibitem{Hansen77} J. P. Hansen, and J.\ J. Weis, Mol. Phys. \textbf{33},
1379 (1977).

\bibitem{Rossky80} P. J. Rossky, J. B Dudowicz, B. L. Tembe, and H. L.
Friedman, J. Chem. Phys. \textbf{73,} 3372 (1980).

\bibitem{Ciccariello82} S. Ciccariello, and D. Gazzillo, J. Chem. Phys. 
\textbf{76,} 1181 (1982).

\bibitem{Pellicane04} G. Pellicane, D. Costa, and C. Caccamo, J. Phys. Chem.
B \textbf{108,} 7538 (2004).

\bibitem{deVries04} R. de Vries, J. Chem. Phys. \textbf{120,} 3475 (2004).

\bibitem{Likos01} see e.g. C. Likos, Phys. Reports \textbf{348}, 267 (2001).

\end{thebibliography}
\end{document}